\documentclass[twocolumn,prd,superscriptaddress,preprintnumbers,nofootinbib]{revtex4}[11pt]
\pdfoutput=1
\usepackage{amsmath,amssymb,graphicx}
\graphicspath{{figs/}}
\usepackage{epsf,verbatim}
\usepackage{hyperref}
\usepackage{comment}
\usepackage{color}
\usepackage{slashed}
\usepackage{subfigure}
\usepackage[usenames,dvipsnames]{xcolor}
\usepackage{comment}
\usepackage{mdwlist, paralist}
\usepackage{rotating}
\usepackage{multirow}

\usepackage[utf8]{inputenc}

\newcommand{\eg}{{\it e.g. }}
\newcommand{\ie}{{\it i.e. }}
\newcommand{\Br}{{\mathrm{Br}}}

\def\to{\rightarrow}

\def\TeV{~{\mbox{TeV}}}

\def\fb{~{\mbox{fb}}}
\def\GeV{~{\mbox{GeV}}}
\def\MeV{~{\mbox{MeV}}}
\def\bS{\mathbb{S}}
\def\mL{\mathcal{L}}
\def\hc{\mathrm{h.c.}}

\emergencystretch=\maxdimen
\begin{document}

\title{A Loop-philic Pseudoscalar}

\author{Gang Li}
\thanks{gangli@pku.edu.cn}
\affiliation{Institute of Theoretical Physics $\&$ State Key Laboratory of Nuclear
Physics and Technology, Peking University, Beijing 100871, China}

\author{Ying-nan Mao}
\thanks{maoyingnan@pku.edu.cn}
\affiliation{Institute of Theoretical Physics $\&$ State Key Laboratory of Nuclear
Physics and Technology, Peking University, Beijing 100871, China}

\author{Yi-Lei Tang}
\thanks{tangyilei15@pku.edu.cn}
\affiliation{Center for High Energy Physics, Peking University, Beijing 100871, China}

\author{Chen Zhang}
\thanks{larry@pku.edu.cn}
\affiliation{Institute of Theoretical Physics $\&$ State Key Laboratory of Nuclear
Physics and Technology, Peking University, Beijing 100871, China}

\author{Yang Zhou}
\thanks{yangzhou9103@gmail.com}
\affiliation{Institute of Theoretical Physics $\&$ State Key Laboratory of Nuclear
Physics and Technology, Peking University, Beijing 100871, China}

\author{Shou-hua Zhu}
\thanks{shzhu@pku.edu.cn}
\affiliation{Institute of Theoretical Physics $\&$ State Key Laboratory of Nuclear
Physics and Technology, Peking University, Beijing 100871, China}
\affiliation{Center for High Energy Physics, Peking University, Beijing 100871, China}
\affiliation{Collaborative Innovation Center of Quantum Matter, Beijing 100871, China}

\begin{abstract}
We construct a weakly-coupled renormalizable model to explain the $750\GeV$ diphoton excess.
The $750\GeV$ resonance (denoted as $X(750)$) is interpreted as a pseudoscalar coming
from a complex singlet. The model also naturally provides a dark matter candidate. One most
attractive feature of the model is that decays of $X(750)$ are all loop-induced so the
diphoton rate is not diluted by unwanted tree level branching fractions. Relevant Yukawa
interactions need not to be tuned to near non-perturbative region to explain the rate.
The model is highly predictive, including the pseudoscalar nature of $X(750)$, and two
nearly mass-degenerate exotic quarks carrying electric charge $5/3$ and $2/3$, respectively.
Rich phenomenology is expected with respect to collider searches, flavor physics and dark
matter detection, if $X(750)$ can be pinned down by future LHC experiments.
\end{abstract}

\maketitle

\setcounter{equation}{0} \setcounter{footnote}{0}

\section{Introduction}

The discovery of the $125\GeV$ Higgs boson~\cite{Aad:2012tfa,Chatrchyan:2012ufa} and the
non-discovery of any new physics signature at the LHC Run I
mark an an amazing triumph of the standard model (SM) and have significant implications for
beyond the SM (BSM) physics searches. Now the guidelines on BSM searches
heavily rely on either highly-debated theoretical issues (\eg Higgs mass fine-tuning), or
observational clues (\eg dark matter (DM)) which
however need not be related to energy scales accessible at present or foreseeable
accelerators. Therefore it is very fortunate if the already obtained LHC Run II
data can reveal some new phenomena which will in turn fast track our BSM searches and
understanding of those fundamental theoretical and observational questions. Recently the
ATLAS and CMS collaborations have reported an excess of diphoton events around $750\GeV$
invariant mass with a local significance of $3.6\sigma$ and $2.6\sigma$, respectively,
using the $13\TeV$ LHC data~\cite{ATLAS:2015diph,CMS:2015dxe}. If this is not merely due to
statistical fluctuations or some unknown systematic uncertainties, the excess will
definitely have far-reaching consequences for any BSM theory and profoundly
shape our understanding of elementary particles and the universe. At present no definite
conclusion can be drawn with regard to the very existence of the resonance, and more data
and refined photon energy calibration are warranted to arrive at a final confirmation or
exclusion. It is nevertheless tempting for theorists to dig out the agents behind the
scene which contribute to such a surprising phenomenon, assuming the existence of the
$750\GeV$ resonance (denoted as $X(750)$). Therefore, the excess triggered a huge amount
of theoretical and phenomenological investigations~\cite{Harigaya:2015ezk,Mambrini:2015wyu,
Backovic:2015fnp,Angelescu:2015uiz,Nakai:2015ptz,Knapen:2015dap,Buttazzo:2015txu,
Pilaftsis:2015ycr,Franceschini:2015kwy,DiChiara:2015vdm,Higaki:2015jag,McDermott:2015sck,
Ellis:2015oso,Low:2015qep,Bellazzini:2015nxw,Gupta:2015zzs,Petersson:2015mkr,
Molinaro:2015cwg,Dutta:2015wqh,Cao:2015pto,Matsuzaki:2015che,Kobakhidze:2015ldh,
Martinez:2015kmn,Cox:2015ckc,Becirevic:2015fmu,No:2015bsn,Demidov:2015zqn,Chao:2015ttq,
Fichet:2015vvy,Curtin:2015jcv,Bian:2015kjt,Chakrabortty:2015hff,Ahmed:2015uqt,
Agrawal:2015dbf,Csaki:2015vek,Falkowski:2015swt,Aloni:2015mxa,Bai:2015nbs,Gabrielli:2015dhk,
Benbrik:2015fyz,Kim:2015ron,Alves:2015jgx,Megias:2015ory,Carpenter:2015ucu,Bernon:2015abk,
Chao:2015nsm,Arun:2015ubr,Han:2015cty,Chang:2015bzc,Chakraborty:2015jvs,Ding:2015rxx,
Han:2015dlp,Han:2015qqj,Luo:2015yio,Chang:2015sdy,Bardhan:2015hcr,Feng:2015wil,
Antipin:2015kgh,Wang:2015kuj,Cao:2015twy,Huang:2015evq,Liao:2015tow,Heckman:2015kqk,
Dhuria:2015ufo,Bi:2015uqd,Kim:2015ksf,Berthier:2015vbb,Cho:2015nxy,Cline:2015msi,
Bauer:2015boy,Chala:2015cev,Barducci:2015gtd,Boucenna:2015pav,Murphy:2015kag,
Hernandez:2015ywg,Dey:2015bur,Pelaggi:2015knk,deBlas:2015hlv,Belyaev:2015hgo,Dev:2015isx,
Huang:2015rkj,Moretti:2015pbj,Patel:2015ulo,Badziak:2015zez,Chakraborty:2015gyj,Cao:2015xjz,
Altmannshofer:2015xfo,Cvetic:2015vit,Gu:2015lxj,Allanach:2015ixl,Davoudiasl:2015cuo,
Craig:2015lra,Das:2015enc,Cheung:2015cug,Liu:2015yec,Zhang:2015uuo,Casas:2015blx,
Hall:2015xds}. The excess is to some extent surprising because it is somewhat
challenging to interpret it naturally in a weakly-coupled model. Taking into account $8\TeV$
results~\cite{Aad:2014ioa,Aad:2015mna,Khachatryan:2015qba,CMS:2015cwa}, the production
mechanism of $X(750)$ is very likely to be gluon fusion, based on parton luminosity
considerations~\cite{Buttazzo:2015txu}. Here we only consider the case that the diphoton
excess indeed corresponds to a new resonance $X(750)$ which is produced in gluon fusion and
subsequently decays to exactly two photons. We note that there are many other interesting
possibilities of exotic kinematics~\cite{Knapen:2015dap} which we won't pursue in the
following. The Landau-Yang theorem~\cite{Landau:1948kw,Yang:1950rg} and the pursuit of a
renormalizable theory prompt us to consider the possibility that $X(750)$ is spin-zero and
both $gg\to X(750)$ and $X(750)\to\gamma\gamma$ proceed via loops. It has been
shown~\cite{Knapen:2015dap} that in this case particles beyond the SM have to be introduced
in the loops so as to provide a sufficient diphoton rate and at the same time to make the
theory compatible with collider constraints from other channels. Even with this addition,
generically it is still difficult to produce the diphoton cross section required to explain
the experimental results (about 5 fb is needed~\cite{Knapen:2015dap,Buttazzo:2015txu})
naturally in renormalizable models. Relevant Yukawa interactions are often tuned to near
non-perturbative region (or a large number of new particles are introduced), and ad hoc assumptions are often made (explicitly or implicitly)
to suppress the tree level decay of $X(750)$ to SM particles. Aimed at tackling these
difficulties, in this Letter we propose a simple, weakly-coupled extension of the SM to
explain the diphoton excess. The model is fully gauge invariant under SM gauge groups and
completely renormalizable. One very attractive feature of the model is that decays of
$X(750)$ are always loop-induced, in this way the early appearance of the diphoton excess
is naturally explained. In the next section we present the contents of the model and
discuss its important phenomenological aspects. Finally we present our discussion and
conclusion.

\section{The Model}

We introduce a complex singlet scalar $\bS$, an inert doublet scalar $\Phi$ and an inert
vector-like doublet quark $Q$, in addition to SM fields. The complete list of relevant
representations and quantum numbers are listed in Table \nolinebreak ~\ref{table:fields}.
CP conservation is required in the extended sector, for both the Lagrangian and the
vacuum. We call $\Phi$ and $Q$ inert because they are odd under a $Z_2$ symmetry, i.e.
$\Phi\to -\Phi,Q\to -Q$ while SM fields and $\bS$ are even. We identify the particle
corresponding to the CP-odd component of $\bS$ as the $X(750)$. In this construction the
particle participating in $gg\to X(750)$ and $X(750)\to\gamma\gamma$ triangle loops
is the inert vector-like quark
$Q={\left(\begin{array}{c}Q_{5/3}\\Q_{2/3}\end{array}\right)}$, where the subscripts
label the electric charge of the weak isospin components. The role of vector-like quarks
in (extra) Higgs boson production and decay has been extensively studied, \eg
~\cite{Bonne:2012im,Moreau:2012da,Ellis:2014dza,Angelescu:2015kga,Alves:2015vob}. Here
the representations of vector-like quarks are chosen carefully so as to produce a large
diphoton rate more easily. The particlular, simple representation of $Q$ listed in
Table ~\ref{table:fields} was noticed by ~\cite{Knapen:2015dap,Chang:2015bzc} and
found to generate relatively large diphoton rate, due to the large multiplicity and
electric charge involved. We note that when making the above statement there is a
crucial idealized assumption, \ie only the di-gluon decay channel dominates the total
width of $X(750)$. If $X(750)$ is a CP-even scalar, its tree level decays
to $W,Z,h,t$ ($h$ denotes the $125\GeV$ Higgs boson) are in general open, albeit they
might be suppressed by mixing angles in certain model construction cases. A CP-odd
scalar is more desirable because it has no tree level trilinear coupling to $WW,ZZ,hh$
(also its loop function has larger asymptotic value). This allures us to also forbid
its tree level coupling to SM fermions (and to $Zh$), which is not possible in a usual
two-Higgs-doublet model setup. This impossibility arises from the particular
representation assignment of the extra Higgs field, which can otherwise be dissolved
by demanding the CP-odd scalar come from a complex singlet, which is just the $\bS$
introduced in our model. In this way the di-gluon channel can naturally dominate the
total width of $X(750)$, without being diluted by unwanted tree level decays.
\begin{table}[t!]
\begin{tabular}{|c|c|c|c|c|c|}
\hline
Field & Spin & $SU(3)_c$ & $SU(2)_L$ & $U(1)_Y$ & $Z_2$ \\
\hline
$\mathbb{S}$  & 0 & \textbf{1} & \textbf{1} & 0 & even \\
\hline
$\Phi$ & 0 & \textbf{1} & \textbf{2} & 1/2 & odd \\
\hline
$Q_L$ & 1/2 & \textbf{3} & \textbf{2} & 7/6 & odd \\
\hline
$Q_R$ & 1/2 & \textbf{3} & \textbf{2} & 7/6 & odd \\
\hline
\end{tabular}
\caption{Field contents in addition to SM in our model. $\bS$ is complex.
\label{table:fields}}
\end{table}
The introduction of $\Phi$ and making $\Phi$ and $Q$ inert is crucial as well. This
becomes evident when we consider the decay of $Q$. On one hand, we would like $Q$
to decay in some manner because stable colored and charged particles are
stringently constrained~\cite{Agashe:2014kda}. On the other hand, if $Q$ is allowed
to decay via mixing with SM quarks as was considered in ~\cite{Knapen:2015dap,
Chang:2015bzc}, the same mixing effect will reintroduce tree level couplings of
$X(750)$ to SM fermions and thus spoil our original goal. Inspired by flavored DM
constructions~\cite{Agrawal:2011ze,Kilic:2015vka,Chao:2015ttq}, we are
led to make $Q$ decay to final states involving DM and thus $\Phi$ and the $Z_2$
assignments are introduced. The lightest particle from the $Z_2$-odd sector, if
color and electrically neutral, becomes a DM candidate, which is taken to be the
CP-even or CP-odd neutral particle from $\Phi$. It is interesting to note that if
the DM particle is heavier than half of the mass of $X(750)$ (which will be
assumed in the following), then $X(750)$ will not have any tree level decay
(even 3-body and multi-body tree level decays are forbidden as well). In such a
case $X(750)$ is therefore called \emph{loop-philic}. In general, loop-induced
decay of $X(750)$ to di-gluon will dominate the width. Minor contributions of
$\gamma\gamma,WW,\gamma Z,ZZ,hh,Zh,tt$ decay modes and decays to final states
involving the additional CP-even $Z_2$-even Higgs boson $h'$ are also expected,
all of which still have to proceed via loops. The diphoton branching ratio is
not diluted by unwanted tree level branching fractions, which is a very
attractive feature of the model.

With all ingredients at our hand, we can now write down the most general
CP-conserving renormalizable gauge-invariant Lagrangian containing the introduced
fields satisfying symmetry assignments dictated by \linebreak[4]
Table ~\ref{table:fields} ($H$ denotes the original Higgs doublet introduced in
the SM)
\begin{eqnarray}
& & \mL=\mL_{HS\Phi}+\mL_{QMass}+\mL_{QGauge}+\mL_{QS}+\mL_{Q\Phi}, \\
& & \mL_{HS\Phi}=(D^{\mu}H)^{\dagger}(D_{\mu}H)+(\partial^{\mu}\bS)^{\dagger}
(\partial_{\mu}\bS) \nonumber \\
& & +(D^{\mu}\Phi)^{\dagger}(D_{\mu}\Phi)-V(H,\bS,\Phi), \label{eqn:HSPhi} \\
& & \mL_{QMass}=-M\bar{Q}_{L}Q_R+\hc, \\
& & \mL_{QGauge}=\bar{Q}_{L}\slashed{D}Q_L+\bar{Q}_{R}\slashed{D}Q_R, \\
& & \mL_{QS}=-\lambda_{QS1}\bS\bar{Q}_{L}Q_R-\lambda_{QS2}\bS^{\dagger}
\bar{Q}_{L}Q_R+\hc, \\
& & \mL_{Q\Phi}=-\lambda_{Q\Phi i}\bar{Q}_L\cdot\Phi u_{Ri}+\hc
\label{eqn:LQPhi}
\end{eqnarray}
In Eq.~\eqref{eqn:LQPhi} $u_{Ri},i=1,2,3$ denotes the three generations of SM up
type quarks and $i$ is summed over. Covariant derivatives in the above equations
are understood to be consistent with the representations and quantum numbers
listed in Table ~\ref{table:fields}. For completeness, we also explicitly write
down the scalar potential $V(H,\bS,\Phi)$ introduced in Eq.~\eqref{eqn:HSPhi}
\begin{eqnarray}
& & V(H,\bS,\Phi)=\mu_{H}^{2}H^{\dagger}H+\lambda_{H}(H^{\dagger}H)^2
+\mu_{S1}^3\bS \nonumber \\
& & +\mu_{S2}^2\bS^2+\mu_{S3}^2\bS^{\dagger}\bS+\mu_{S4}\bS^3
+\mu_{S5}(\bS^{\dagger}\bS)\bS+\lambda_{S1}\bS^4 \nonumber \\
& & +\lambda_{S2}(\bS^{\dagger}\bS)\bS^2+\lambda_{S3}(\bS^{\dagger}\bS)^2
+\mu_{\Phi}^2\Phi^{\dagger}\Phi+\lambda_{\Phi}(\Phi^{\dagger}\Phi)^2
\nonumber \\
& & +\mu_{HS}(H^{\dagger}H)\bS+\lambda_{HS1}(H^{\dagger}H)\bS^2 \nonumber \\
& & +\lambda_{HS2}(H^{\dagger}H)(\bS^{\dagger}\bS)
+\lambda_{H\Phi 1}(H^{\dagger}H)(\Phi^{\dagger}\Phi) \nonumber \\
& & +\lambda_{H\Phi 2}(H^{\dagger}\Phi)(\Phi^{\dagger}H)
+\lambda_{H\Phi 3}(H^{\dagger}\Phi)^2
\nonumber \\
& & +\mu_{S\Phi}\bS(\Phi^{\dagger}\Phi)
+\lambda_{S\Phi 1}\bS^2(\Phi^{\dagger}\Phi)
+\lambda_{S\Phi 2}(\bS^{\dagger}\bS)(\Phi^{\dagger}\Phi) \nonumber \\
& & +\hc, \label{eqn:pot}
\end{eqnarray}
In Eq.~\eqref{eqn:pot} $\hc$ in the last line represents the hermitian conjugate
of the terms which are present in the potential but are not self-conjugate. We
require all $\mu's,\lambda's$ to be real so as to make the Lagrangian
CP-invariant ($M$ can always be made real by rephasing $Q_L,Q_R$ fields). We
assume $Z_2$ is also preserved by vacuum and thus $<\Phi>=0$. It is legitimate
to shift $\bS$ in order to make $<\bS>=0$, which will be assumed in the
following. This offers the convenience that the tree level mass of $Q$ is just
$M$. The mass degeneracy of $Q_{5/3}$ and $Q_{2/3}$ is broken only by loop
effects by which gauge boson loops will make $Q_{5/3}$ slightly heavier than
$Q_{2/3}$. We do not expect loops induced by $\mL_{Q\Phi}$ to substantially
enlarge this mass difference. Therefore in the following calculation we simply
take $Q_{5/3}$ and $Q_{2/3}$ to be mass-degenerate at $M$.

\begin{figure}[ht]
\includegraphics[width=2.2in]{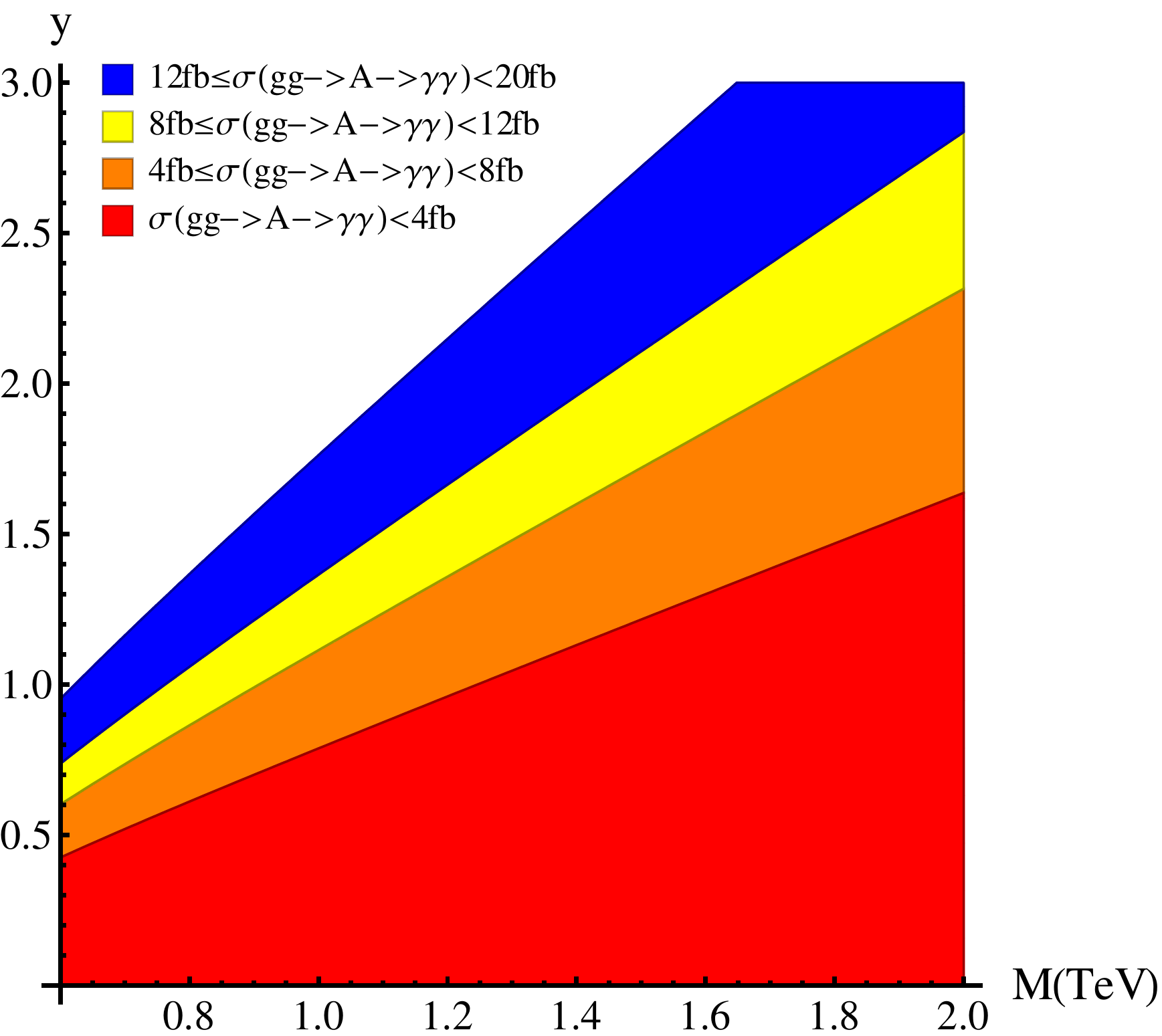}
\caption{Diphoton rate in our model plotted in the $M-y$ plane. The color
scheme is described in the plot.\label{fig:rate}}
\end{figure}

For notational convenience, we write $\bS=\frac{1}{\sqrt{2}} (S+iA)$ in which
the particle excitation of $A$ is just $X(750)$, with mass $m_A=750\GeV$. To
discuss the diphoton rate of $A$, the most important part of the Lagrangian is
$\mL_{QS}$, which gives the following effective Lagrangian
\begin{eqnarray}
& & \mL_{AQQ}=-y(\bar{Q}_{5/3}i\gamma^{5}Q_{5/3}+\bar{Q}_{2/3}i\gamma^{5}Q_{2/3})A,
\end{eqnarray}
where we introduced the effective Yukawa coupling constant for $AQQ$ interaction
$y\equiv\frac{\lambda_{QS1}-\lambda_{QS2}}{\sqrt{2}}$. We also introduce an
effective Yukawa coupling constant $y'$ for $h'QQ$ interaction
$y'\equiv\frac{\lambda_{QS1}+\lambda_{QS2}}{\sqrt{2}}$. Assuming
$x\equiv\frac{m_A^2}{4M^2}<1$, the leading order partial
widths of $A\to gg,\gamma\gamma$ are calculated as~\cite{Djouadi:2005gi,
Djouadi:2005gj}
\begin{eqnarray}
& & \Gamma(A\to gg)=\frac{\alpha_s^2 m_A^3y^2}{8\pi^3 M^2}
\Big(\frac{\arcsin^2\sqrt{x}}{x}\Big)^2, \\
& & \Gamma(A\to\gamma\gamma)=\frac{841\alpha_{em}^2 m_A^3y^2}{576\pi^3 M^2}
\Big(\frac{\arcsin^2\sqrt{x}}{x}\Big)^2,
\end{eqnarray}
For a loop-philic $A$, other important decay channels considered here include
$WW,ZZ,\gamma Z$, with approximate partial width ratios calculated to be
$\frac{\Gamma(A\to WW)}{\Gamma(A\to\gamma\gamma)}=0.91,
\frac{\Gamma(A\to ZZ)}{\Gamma(A\to\gamma\gamma)}=0.60,
\frac{\Gamma(A\to\gamma Z)}{\Gamma(A\to\gamma\gamma)}=0.06$. Loop-induced
decays to $tt,cc,uu,tc,tu,cu$ and $hh,h'h',hh',Zh,Zh'$ are expected to be
small if we require $\lambda_{Q\Phi i},y'$ and the mixing in the CP-even Higgs
sector to be small. Therefore for total width calculation, we only take
into account $A\to gg,\gamma\gamma,WW,ZZ,\gamma Z$. All relevant K-factors are
taken to be $1$, especially because it is expected that for $gg$ initial and
final states, the K-factor effect cancels a lot when calculating the rate.
MSTW2008 PDF~\cite{Martin:2009iq} is used for parton-parton luminosity
calculations. In \linebreak[4] FIG. ~\ref{fig:rate} we color the parameter
region where the diphoton $\sigma\times\Br$ can reach $4,8,12,20\fb$. We note
that for the heavy DM case ($\gtrsim 550\GeV$) (which is one option favored by
relic density considerations, see the paragraph on DM below), essentially
there is no lower bound on the mass of $Q_{2/3}$ other than the DM mass
assuming $100\%$ decay to top final states~\cite{Chao:2015ttq}.
~\footnote{There are collider searches for top-quark partners with $5/3$
electric charge~\cite{Chatrchyan:2013wfa,Aad:2015mba,Aad:2015gdg,CMS:2015alb}.
However their bounds cannot be directly applied to our case because some
different kinematic features (\eg $H_T$ distribution) are involved.}
Therefore for
a wide range of $M$, there exists a fully perturbative region of $y$ to
realize the observed diphoton rate. If we take a benchmark point
$M=1\TeV,y=1.0$, we will get $\sigma(gg\to A\to\gamma\gamma)=6.4\fb$, which
just falls in the right range needed to explain the excess. The total width of
$A$ at this benchmark point is about $18\MeV$ which is not able to also
account for the apparently large width hinted by ATLAS~\cite{ATLAS:2015diph}.
However the hint is quite preliminary and not supported by
CMS~\cite{CMS:2015dxe} so we still stick to the narrow width interpretation.
With the above-mentioned partial width ratios, it is easy to check that for
$X(750)$ no bounds are relevant at present with regard to LHC searches in
dijet,$WW,ZZ,\gamma Z$ channels~\cite{CMS:2015neg,CMS:2015lda,Aad:2015agg,
CMS:2013ada,Aad:2015kna,CMS:2015mda,Aad:2014fha}, which in turn provides a
natural reason why the diphoton channel popped up first. We have checked that
the modification of the $gg\to h\to\gamma\gamma$ signal strength due to $Q$
is safely negligible. The contribution of $Q_{5/3}$ and $Q_{2/3}$ to
electroweak precision observables is also negligible due to their
vector-like nature and approximate mass-degeneracy~\cite{He:2001tp}.

A few remarks are in order here with respect to DM phenomenology in the
model. In fact, the model possesses a decoupling limit in which makes DM
phenomenology here approach the usual inert doublet model
(IDM)~\cite{Deshpande:1977rw}. There are three additional factors
contributing to DM phenomenology compared with IDM. The first is the
mixing of two CP-even neutral Higgs bosons in the $Z_2$-even sector.
The second is additional trilinear and quartic couplings in the scalar
sector. The third is the new $\mL_{Q\Phi}$ interactions. The impact of
all three factors can be made small by keeping relevant mixing and
coupling small and thus the model approaches the IDM limit. It is
known that the IDM has a region of parameter space corresponding to
large DM mass ($\gtrsim 550\GeV$) which can give correct relic density
and satisfy direct detection bound~\cite{Krawczyk:2015xhl}. Thus our
model is expected to be able to produce correct DM relic density and
be safe from direct detection constraints. We note that it is
interesting, though beyond the scope of this Letter, to consider the
case in which one or more of these three factors make a sizable
contribution to DM annihilation or direct detection and thus
have interplay with ordinary IDM.

Our model can be tested in the future in various manners besides the
diphoton and diboson rates. First, our model predicts a CP-odd scalar
whose CP-property can be tested via investigating the final state
differential distribution of $gg\to A\to WW,ZZ,\gamma Z$. Second, our
model predicts an additional CP-even scalar $h'$ which can mix with
the $125\GeV$ Higgs boson. Its trail can be detected in future
precision electroweak or Higgs studies. This additional CP-even
scalar can also be produced directly at colliders and decay to SM
final states. Inert vector-like quarks might contribute in its
production and decay loop, however the associated rate is not linked
tightly to $X(750)$ production because in our construction $y'$ can
be adjusted independently with respect to $y$. Third, exotic quarks of
electric charge $5/3$ and $2/3$ are expected and nearly
mass-degenerate, which can be searched for at future hadron or linear
colliders, via signatures such as 
$\text{lepton(s)}+\text{jets}+\slashed{E}_T$ (leptons and jets come 
from $t,W,Z$ from $Q_{5/3},Q_{2/3}$ decay and can often be soft). 
The chirality structure of $\mL_{Q\Phi}$ can be further studied via 
studying the polarization of the top quark in the process, \eg 
in the same spirit as ~\cite{Berger:2012an}. $\mL_{Q\Phi}$ also 
contributes to flavor observable such as $\Delta m_D$, which at 
present provides the constraint
$|\lambda_{Q\Phi u}\lambda_{Q\Phi c}|\lesssim 5\times 10^{-3}$.
More parameter space on these two couplings is expected to be probed
by future improved flavor experiments and refined lattice
calculations. Furthermore, scalars from the inert doublet in our
model can be probed by future dark matter experiments and production
at very high energy hadron, lepton, and lepton-hadron colliders.

\section{Discussion and Conclusion}

In this Letter we introduced a weakly-coupled renormalizable model in which
there exists a particle whose tree level decays are all forbidden and therefore
can only decay via quantum effects. This particle is a pseudoscalar coming from
a complex singlet scalar field. With appropriate vector-like fermions conspiring
in the loop it is easy to produce a sufficiently large diphoton rate for $X(750)$
without driving relevant Yukawa couplings to non-perturbative region.
Interestingly enough, the prohibition of tree level decays of $X(750)$ naturally
leads to a dark matter candidate in the model. If $X(750)$ can be confirmed in
the future, there should be rich phenomenology with respect to collider searches,
flavor physics and dark matter detection. We note that this loop-philic
construction is not specific to the $(\textbf{3},\textbf{2},7/6)$ representation
and quantum number chosen for $Q$. Other representations and quantum numbers
(such as those considered in ~\cite{Knapen:2015dap}) can be easily accommodated
as long as appropriate representations for inert scalars are chosen, although
one might be confronted with different (perhaps more stringent) theoretical and
phenomenological constraints in such cases.

The $750\GeV$ diphoton excess is a huge surprise to the high energy physics
community. On one hand, physics beyond the SM has been sought for for a long time
at colliders without success. The diphoton excess, though preliminary, has an
unexpected possibility to become the first smoking gun signature of physics
beyond the SM at the LHC, and thus clearly warrants further experimental and
theoretical investigations. On the other hand, there have been two major trends
in new physics model building, \ie weak dynamics (\eg supersymmetry) and strong
dynamics (\eg composite Higgs). Weak dynamics has some special advantages such as
its renormalizability and being easier to make predictions. However the diphoton
excess, at first glance, is difficult to explain naturally in a completely
weakly-coupled theory. Our study shows that with appropriate model construction
it is possible to naturally accommodate the diphoton excess with the addition of
only a few particles in a completely weakly-coupled renormalizable
gauge-invariant framework. In this Letter a minimal realization of such a
framework is presented, in which the diphoton excess appears the earliest is
fully expected due to the loop-philic nature of $X(750)$.

\vfil

\subsection*{Acknowledgements}

We would like to thank Yan-Dong Liu and Jia-Shu Lu for helpful discussions.
This work was supported in part by the Natural Science Foundation of China
(Grants No. 11135003 and No. 11375014).


\bibliography{diphoton_v4}

\begin{thebibliography}{100}

\bibitem{Aad:2012tfa}
ATLAS, G.~Aad {\em et~al.},
\newblock Phys. Lett. {\bf B716}, 1 (2012), 1207.7214.

\bibitem{Chatrchyan:2012ufa}
CMS, S.~Chatrchyan {\em et~al.},
\newblock Phys.Lett. {\bf B716}, 30 (2012), 1207.7235.

\bibitem{ATLAS:2015diph}
ATLAS,
\newblock (2015), ATLAS-CONF-2015-081.

\bibitem{CMS:2015dxe}
CMS,
\newblock (2015), CMS-PAS-EXO-15-004.

\bibitem{Harigaya:2015ezk}
K.~Harigaya and Y.~Nomura,
\newblock (2015), 1512.04850.

\bibitem{Mambrini:2015wyu}
Y.~Mambrini, G.~Arcadi, and A.~Djouadi,
\newblock (2015), 1512.04913.

\bibitem{Backovic:2015fnp}
M.~Backovic, A.~Mariotti, and D.~Redigolo,
\newblock (2015), 1512.04917.

\bibitem{Angelescu:2015uiz}
A.~Angelescu, A.~Djouadi, and G.~Moreau,
\newblock (2015), 1512.04921.

\bibitem{Nakai:2015ptz}
Y.~Nakai, R.~Sato, and K.~Tobioka,
\newblock (2015), 1512.04924.

\bibitem{Knapen:2015dap}
S.~Knapen, T.~Melia, M.~Papucci, and K.~Zurek,
\newblock (2015), 1512.04928.

\bibitem{Buttazzo:2015txu}
D.~Buttazzo, A.~Greljo, and D.~Marzocca,
\newblock (2015), 1512.04929.

\bibitem{Pilaftsis:2015ycr}
A.~Pilaftsis,
\newblock (2015), 1512.04931.

\bibitem{Franceschini:2015kwy}
R.~Franceschini {\em et~al.},
\newblock (2015), 1512.04933.

\bibitem{DiChiara:2015vdm}
S.~Di~Chiara, L.~Marzola, and M.~Raidal,
\newblock (2015), 1512.04939.

\bibitem{Higaki:2015jag}
T.~Higaki, K.~S. Jeong, N.~Kitajima, and F.~Takahashi,
\newblock (2015), 1512.05295.

\bibitem{McDermott:2015sck}
S.~D. McDermott, P.~Meade, and H.~Ramani,
\newblock (2015), 1512.05326.

\bibitem{Ellis:2015oso}
J.~Ellis, S.~A.~R. Ellis, J.~Quevillon, V.~Sanz, and T.~You,
\newblock (2015), 1512.05327.

\bibitem{Low:2015qep}
M.~Low, A.~Tesi, and L.-T. Wang,
\newblock (2015), 1512.05328.

\bibitem{Bellazzini:2015nxw}
B.~Bellazzini, R.~Franceschini, F.~Sala, and J.~Serra,
\newblock (2015), 1512.05330.

\bibitem{Gupta:2015zzs}
R.~S. Gupta, S.~Jager, Y.~Kats, G.~Perez, and E.~Stamou,
\newblock (2015), 1512.05332.

\bibitem{Petersson:2015mkr}
C.~Petersson and R.~Torre,
\newblock (2015), 1512.05333.

\bibitem{Molinaro:2015cwg}
E.~Molinaro, F.~Sannino, and N.~Vignaroli,
\newblock (2015), 1512.05334.

\bibitem{Dutta:2015wqh}
B.~Dutta, Y.~Gao, T.~Ghosh, I.~Gogoladze, and T.~Li,
\newblock (2015), 1512.05439.

\bibitem{Cao:2015pto}
Q.-H. Cao, Y.~Liu, K.-P. Xie, B.~Yan, and D.-M. Zhang,
\newblock (2015), 1512.05542.

\bibitem{Matsuzaki:2015che}
S.~Matsuzaki and K.~Yamawaki,
\newblock (2015), 1512.05564.

\bibitem{Kobakhidze:2015ldh}
A.~Kobakhidze, F.~Wang, L.~Wu, J.~M. Yang, and M.~Zhang,
\newblock (2015), 1512.05585.

\bibitem{Martinez:2015kmn}
R.~Martinez, F.~Ochoa, and C.~F. Sierra,
\newblock (2015), 1512.05617.

\bibitem{Cox:2015ckc}
P.~Cox, A.~D. Medina, T.~S. Ray, and A.~Spray,
\newblock (2015), 1512.05618.

\bibitem{Becirevic:2015fmu}
D.~Becirevic, E.~Bertuzzo, O.~Sumensari, and R.~Z. Funchal,
\newblock (2015), 1512.05623.

\bibitem{No:2015bsn}
J.~M. No, V.~Sanz, and J.~Setford,
\newblock (2015), 1512.05700.

\bibitem{Demidov:2015zqn}
S.~V. Demidov and D.~S. Gorbunov,
\newblock (2015), 1512.05723.

\bibitem{Chao:2015ttq}
W.~Chao, R.~Huo, and J.-H. Yu,
\newblock (2015), 1512.05738.

\bibitem{Fichet:2015vvy}
S.~Fichet, G.~von Gersdorff, and C.~Royon,
\newblock (2015), 1512.05751.

\bibitem{Curtin:2015jcv}
D.~Curtin and C.~B. Verhaaren,
\newblock (2015), 1512.05753.

\bibitem{Bian:2015kjt}
L.~Bian, N.~Chen, D.~Liu, and J.~Shu,
\newblock (2015), 1512.05759.

\bibitem{Chakrabortty:2015hff}
J.~Chakrabortty, A.~Choudhury, P.~Ghosh, S.~Mondal, and T.~Srivastava,
\newblock (2015), 1512.05767.

\bibitem{Ahmed:2015uqt}
A.~Ahmed, B.~M. Dillon, B.~Grzadkowski, J.~F. Gunion, and Y.~Jiang,
\newblock (2015), 1512.05771.

\bibitem{Agrawal:2015dbf}
P.~Agrawal, J.~Fan, B.~Heidenreich, M.~Reece, and M.~Strassler,
\newblock (2015), 1512.05775.

\bibitem{Csaki:2015vek}
C.~Csaki, J.~Hubisz, and J.~Terning,
\newblock (2015), 1512.05776.

\bibitem{Falkowski:2015swt}
A.~Falkowski, O.~Slone, and T.~Volansky,
\newblock (2015), 1512.05777.

\bibitem{Aloni:2015mxa}
D.~Aloni, K.~Blum, A.~Dery, A.~Efrati, and Y.~Nir,
\newblock (2015), 1512.05778.

\bibitem{Bai:2015nbs}
Y.~Bai, J.~Berger, and R.~Lu,
\newblock (2015), 1512.05779.

\bibitem{Gabrielli:2015dhk}
E.~Gabrielli {\em et~al.},
\newblock (2015), 1512.05961.

\bibitem{Benbrik:2015fyz}
R.~Benbrik, C.-H. Chen, and T.~Nomura,
\newblock (2015), 1512.06028.

\bibitem{Kim:2015ron}
J.~S. Kim, J.~Reuter, K.~Rolbiecki, and R.~R. de~Austri,
\newblock (2015), 1512.06083.

\bibitem{Alves:2015jgx}
A.~Alves, A.~G. Dias, and K.~Sinha,
\newblock (2015), 1512.06091.

\bibitem{Megias:2015ory}
E.~Megias, O.~Pujolas, and M.~Quiros,
\newblock (2015), 1512.06106.

\bibitem{Carpenter:2015ucu}
L.~M. Carpenter, R.~Colburn, and J.~Goodman,
\newblock (2015), 1512.06107.

\bibitem{Bernon:2015abk}
J.~Bernon and C.~Smith,
\newblock (2015), 1512.06113.

\bibitem{Chao:2015nsm}
W.~Chao,
\newblock (2015), 1512.06297.

\bibitem{Arun:2015ubr}
M.~T. Arun and P.~Saha,
\newblock (2015), 1512.06335.

\bibitem{Han:2015cty}
C.~Han, H.~M. Lee, M.~Park, and V.~Sanz,
\newblock (2015), 1512.06376.

\bibitem{Chang:2015bzc}
S.~Chang,
\newblock (2015), 1512.06426.

\bibitem{Chakraborty:2015jvs}
I.~Chakraborty and A.~Kundu,
\newblock (2015), 1512.06508.

\bibitem{Ding:2015rxx}
R.~Ding, L.~Huang, T.~Li, and B.~Zhu,
\newblock (2015), 1512.06560.

\bibitem{Han:2015dlp}
H.~Han, S.~Wang, and S.~Zheng,
\newblock (2015), 1512.06562.

\bibitem{Han:2015qqj}
X.-F. Han and L.~Wang,
\newblock (2015), 1512.06587.

\bibitem{Luo:2015yio}
M.-x. Luo, K.~Wang, T.~Xu, L.~Zhang, and G.~Zhu,
\newblock (2015), 1512.06670.

\bibitem{Chang:2015sdy}
J.~Chang, K.~Cheung, and C.-T. Lu,
\newblock (2015), 1512.06671.

\bibitem{Bardhan:2015hcr}
D.~Bardhan {\em et~al.},
\newblock (2015), 1512.06674.

\bibitem{Feng:2015wil}
T.-F. Feng, X.-Q. Li, H.-B. Zhang, and S.-M. Zhao,
\newblock (2015), 1512.06696.

\bibitem{Antipin:2015kgh}
O.~Antipin, M.~Mojaza, and F.~Sannino,
\newblock (2015), 1512.06708.

\bibitem{Wang:2015kuj}
F.~Wang, L.~Wu, J.~M. Yang, and M.~Zhang,
\newblock (2015), 1512.06715.

\bibitem{Cao:2015twy}
J.~Cao {\em et~al.},
\newblock (2015), 1512.06728.

\bibitem{Huang:2015evq}
F.~P. Huang, C.~S. Li, Z.~L. Liu, and Y.~Wang,
\newblock (2015), 1512.06732.

\bibitem{Liao:2015tow}
W.~Liao and H.-q. Zheng,
\newblock (2015), 1512.06741.

\bibitem{Heckman:2015kqk}
J.~J. Heckman,
\newblock (2015), 1512.06773.

\bibitem{Dhuria:2015ufo}
M.~Dhuria and G.~Goswami,
\newblock (2015), 1512.06782.

\bibitem{Bi:2015uqd}
X.-J. Bi, Q.-F. Xiang, P.-F. Yin, and Z.-H. Yu,
\newblock (2015), 1512.06787.

\bibitem{Kim:2015ksf}
J.~S. Kim, K.~Rolbiecki, and R.~R. de~Austri,
\newblock (2015), 1512.06797.

\bibitem{Berthier:2015vbb}
L.~Berthier, J.~M. Cline, W.~Shepherd, and M.~Trott,
\newblock (2015), 1512.06799.

\bibitem{Cho:2015nxy}
W.~S. Cho {\em et~al.},
\newblock (2015), 1512.06824.

\bibitem{Cline:2015msi}
J.~M. Cline and Z.~Liu,
\newblock (2015), 1512.06827.

\bibitem{Bauer:2015boy}
M.~Bauer and M.~Neubert,
\newblock (2015), 1512.06828.

\bibitem{Chala:2015cev}
M.~Chala, M.~Duerr, F.~Kahlhoefer, and K.~Schmidt-Hoberg,
\newblock (2015), 1512.06833.

\bibitem{Barducci:2015gtd}
D.~Barducci, A.~Goudelis, S.~Kulkarni, and D.~Sengupta,
\newblock (2015), 1512.06842.

\bibitem{Boucenna:2015pav}
S.~M. Boucenna, S.~Morisi, and A.~Vicente,
\newblock (2015), 1512.06878.

\bibitem{Murphy:2015kag}
C.~W. Murphy,
\newblock (2015), 1512.06976.

\bibitem{Hernandez:2015ywg}
A.~E.~C. Hernandez and I.~Nisandzic,
\newblock (2015), 1512.07165.

\bibitem{Dey:2015bur}
U.~K. Dey, S.~Mohanty, and G.~Tomar,
\newblock (2015), 1512.07212.

\bibitem{Pelaggi:2015knk}
G.~M. Pelaggi, A.~Strumia, and E.~Vigiani,
\newblock (2015), 1512.07225.

\bibitem{deBlas:2015hlv}
J.~de~Blas, J.~Santiago, and R.~Vega-Morales,
\newblock (2015), 1512.07229.

\bibitem{Belyaev:2015hgo}
A.~Belyaev {\em et~al.},
\newblock (2015), 1512.07242.

\bibitem{Dev:2015isx}
P.~S.~B. Dev and D.~Teresi,
\newblock (2015), 1512.07243.

\bibitem{Huang:2015rkj}
W.-C. Huang, Y.-L.~S. Tsai, and T.-C. Yuan,
\newblock (2015), 1512.07268.

\bibitem{Moretti:2015pbj}
S.~Moretti and K.~Yagyu,
\newblock (2015), 1512.07462.

\bibitem{Patel:2015ulo}
K.~M. Patel and P.~Sharma,
\newblock (2015), 1512.07468.

\bibitem{Badziak:2015zez}
M.~Badziak,
\newblock (2015), 1512.07497.

\bibitem{Chakraborty:2015gyj}
S.~Chakraborty, A.~Chakraborty, and S.~Raychaudhuri,
\newblock (2015), 1512.07527.

\bibitem{Cao:2015xjz}
Q.-H. Cao, S.-L. Chen, and P.-H. Gu,
\newblock (2015), 1512.07541.

\bibitem{Altmannshofer:2015xfo}
W.~Altmannshofer {\em et~al.},
\newblock (2015), 1512.07616.

\bibitem{Cvetic:2015vit}
M.~Cvetic, J.~Halverson, and P.~Langacker,
\newblock (2015), 1512.07622.

\bibitem{Gu:2015lxj}
J.~Gu and Z.~Liu,
\newblock (2015), 1512.07624.

\bibitem{Allanach:2015ixl}
B.~C. Allanach, P.~S.~B. Dev, S.~A. Renner, and K.~Sakurai,
\newblock (2015), 1512.07645.

\bibitem{Davoudiasl:2015cuo}
H.~Davoudiasl and C.~Zhang,
\newblock (2015), 1512.07672.

\bibitem{Craig:2015lra}
N.~Craig, P.~Draper, C.~Kilic, and S.~Thomas,
\newblock (2015), 1512.07733.

\bibitem{Das:2015enc}
K.~Das and S.~K. Rai,
\newblock (2015), 1512.07789.

\bibitem{Cheung:2015cug}
K.~Cheung, P.~Ko, J.~S. Lee, J.~Park, and P.-Y. Tseng,
\newblock (2015), 1512.07853.

\bibitem{Liu:2015yec}
J.~Liu, X.-P. Wang, and W.~Xue,
\newblock (2015), 1512.07885.

\bibitem{Zhang:2015uuo}
J.~Zhang and S.~Zhou,
\newblock (2015), 1512.07889.

\bibitem{Casas:2015blx}
J.~A. Casas, J.~R. Espinosa, and J.~M. Moreno,
\newblock (2015), 1512.07895.

\bibitem{Hall:2015xds}
L.~J. Hall, K.~Harigaya, and Y.~Nomura,
\newblock (2015), 1512.07904.

\bibitem{Aad:2014ioa}
ATLAS, G.~Aad {\em et~al.},
\newblock Phys. Rev. Lett. {\bf 113}, 171801 (2014), 1407.6583.

\bibitem{Aad:2015mna}
ATLAS, G.~Aad {\em et~al.},
\newblock Phys. Rev. {\bf D92}, 032004 (2015), 1504.05511.

\bibitem{Khachatryan:2015qba}
CMS, V.~Khachatryan {\em et~al.},
\newblock Phys. Lett. {\bf B750}, 494 (2015), 1506.02301.

\bibitem{CMS:2015cwa}
CMS,
\newblock (2015), CMS-PAS-EXO-12-045.

\bibitem{Landau:1948kw}
L.~D. Landau,
\newblock Dokl. Akad. Nauk Ser. Fiz. {\bf 60}, 207 (1948).

\bibitem{Yang:1950rg}
C.-N. Yang,
\newblock Phys. Rev. {\bf 77}, 242 (1950).

\bibitem{Bonne:2012im}
N.~Bonne and G.~Moreau,
\newblock Phys. Lett. {\bf B717}, 409 (2012), 1206.3360.

\bibitem{Moreau:2012da}
G.~Moreau,
\newblock Phys. Rev. {\bf D87}, 015027 (2013), 1210.3977.

\bibitem{Ellis:2014dza}
S.~A.~R. Ellis, R.~M. Godbole, S.~Gopalakrishna, and J.~D. Wells,
\newblock JHEP {\bf 09}, 130 (2014), 1404.4398.

\bibitem{Angelescu:2015kga}
A.~Angelescu, A.~Djouadi, and G.~Moreau,
\newblock (2015), 1510.07527.

\bibitem{Alves:2015vob}
A.~Alves, D.~A. Camargo, and A.~G. Dias,
\newblock (2015), 1511.04449.

\bibitem{Agashe:2014kda}
Particle Data Group, K.~A. Olive {\em et~al.},
\newblock Chin. Phys. {\bf C38}, 090001 (2014).

\bibitem{Agrawal:2011ze}
P.~Agrawal, S.~Blanchet, Z.~Chacko, and C.~Kilic,
\newblock Phys. Rev. {\bf D86}, 055002 (2012), 1109.3516.

\bibitem{Kilic:2015vka}
C.~Kilic, M.~D. Klimek, and J.-H. Yu,
\newblock Phys. Rev. {\bf D91}, 054036 (2015), 1501.02202.

\bibitem{Djouadi:2005gi}
A.~Djouadi,
\newblock Phys. Rept. {\bf 457}, 1 (2008), hep-ph/0503172.

\bibitem{Djouadi:2005gj}
A.~Djouadi,
\newblock Phys. Rept. {\bf 459}, 1 (2008), hep-ph/0503173.

\bibitem{Martin:2009iq}
A.~D. Martin, W.~J. Stirling, R.~S. Thorne, and G.~Watt,
\newblock Eur. Phys. J. {\bf C63}, 189 (2009), 0901.0002.

\bibitem{Chatrchyan:2013wfa}
CMS, S.~Chatrchyan {\em et~al.},
\newblock Phys. Rev. Lett. {\bf 112}, 171801 (2014), 1312.2391.

\bibitem{Aad:2015mba}
ATLAS, G.~Aad {\em et~al.},
\newblock Phys. Rev. {\bf D91}, 112011 (2015), 1503.05425.

\bibitem{Aad:2015gdg}
ATLAS, G.~Aad {\em et~al.},
\newblock JHEP {\bf 10}, 150 (2015), 1504.04605.

\bibitem{CMS:2015alb}
CMS,
\newblock (2015), CMS-PAS-B2G-15-006.

\bibitem{CMS:2015neg}
CMS,
\newblock (2015), CMS-PAS-EXO-14-005.

\bibitem{CMS:2015lda}
CMS,
\newblock (2015), CMS-PAS-HIG-14-008.

\bibitem{Aad:2015agg}
ATLAS, G.~Aad {\em et~al.},
\newblock (2015), 1509.00389.

\bibitem{CMS:2013ada}
CMS,
\newblock (2013), CMS-PAS-HIG-13-014.

\bibitem{Aad:2015kna}
ATLAS, G.~Aad {\em et~al.},
\newblock (2015), 1507.05930.

\bibitem{CMS:2015mda}
CMS,
\newblock (2015), CMS-PAS-HIG-14-007.

\bibitem{Aad:2014fha}
ATLAS, G.~Aad {\em et~al.},
\newblock Phys. Lett. {\bf B738}, 428 (2014), 1407.8150.

\bibitem{He:2001tp}
H.-J. He, N.~Polonsky, and S.-f. Su,
\newblock Phys. Rev. {\bf D64}, 053004 (2001), hep-ph/0102144.

\bibitem{Deshpande:1977rw}
N.~G. Deshpande and E.~Ma,
\newblock Phys. Rev. {\bf D18}, 2574 (1978).

\bibitem{Krawczyk:2015xhl}
M.~Krawczyk, N.~Darvishi, and D.~Sokolowska,
\newblock {The Inert Doublet Model and its extensions},
\newblock 2015, 1512.06437.

\bibitem{Berger:2012an}
E.~L. Berger, Q.-H. Cao, J.-H. Yu, and H.~Zhang,
\newblock Phys. Rev. Lett. {\bf 109}, 152004 (2012), 1207.1101.

\end{thebibliography}
\bibliographystyle{h-physrev}

\end{document}